\documentclass[journal]{IEEEtran}

\ifCLASSINFOpdf
  \usepackage[pdftex]{graphicx}
\else
\fi

\usepackage{amsmath}
\usepackage{algorithmic}
\usepackage{amsthm}
\usepackage{amssymb}

\usepackage{color}
\usepackage{algorithm}

\usepackage{hyperref}

\usepackage{subcaption}

\begin{document}

\title{On the design of a decision engine for connected vehicles with an application to congestion management}

\author{Rodrigo~Ord\'{o}\~{n}ez-Hurtado,
        Giovanni~Russo,~\IEEEmembership{Member,~IEEE,},
	Sam~Sinnott,
        and~Robert~Shorten,~\IEEEmembership{Member,~IEEE}

\thanks{Rodrigo Ord\'{o}\~{n}ez-Hurtado, Sam~Sinnott and Robert Shorten are with the University College Dublin, School of Electrical, Electronic and Communications Engineering. 
        {\tt\small rodrigo.ordonez-hurtado@ucd.ie}, {\tt\small sam.sinnott@ucdconnect.ie}, {\tt\small robert.shorten@ucd.ie}}%
\thanks{Rodrigo Ord\'{o}\~{n}ez-Hurtado and Giovanni Russo are with IBM Research - Ireland.
         {\tt\small rodrigo.ordonez.hurtado@ibm.com}, {\tt\small grusso@ie.ibm.com}}%
\thanks{Sam~Sinnott was involved in this work while he was doing his M.E. degree in University College Dublin.}}

\markboth{}%
{Shell \MakeLowercase{\textit{et al.}}: On the design of a decision engine for connected vehicles with an application to congestion management}

%

\maketitle

\begin{abstract}
Vehicles are becoming connected entities. As a result, a likely scenario is that such entities might be literally bombarded with information from a multitude of devices. In this context, a key challenging requirement for both connected and autonomous vehicles is that they will need to become {\em cognitive} bodies, able  to parse information and use only the pieces of information that are relevant to the vehicle in the context of a given journey. In order to address this fundamental requirement, a {\em decision engine} is presented in this paper. The engine makes it possible for the vehicle to understand which pieces of information are really relevant, and subsequently to process only those pieces of information. In order to illustrate the key features of our system, we show that it is possible to build upon the engine to develop a distributed traffic management system, and then we validate such a system via both conventional (numerical and SUMO-based) simulations and a Hardware-in-the-Loop (HIL) platform. Both the conventional simulations and the HIL validation showed that the engine can be effectively used to design a distributed traffic management system. 
\end{abstract}


\IEEEpeerreviewmaketitle

\section{Introduction}

\IEEEPARstart{V}{ehicles} are dramatically changing, transitioning from being isolated entities to becoming connected {\em devices}, part of a networked {\em ecosystem} involving both other vehicles (via V2V communication infrastructures) and smart sensors deployed on the wayside infrastructures (via V2I communication). The concept of connected vehicle is showing a great industrial potential: recent studies, for example, predict that by 2018, 17\% of vehicles will be connected to the Internet and 13\% of potential vehicles' buyers are no longer willing to buy a new vehicle without Internet access \cite{Connectivity,McKinseyInternet}.

In this context, over the years, several standards and tools have emerged to enable a networked ecosystem around connected vehicles. For example, with the establishment of the OBD-II standard, the ease of access to OBD-II interfaces, and a boom in the development of smartphone apps, cars became part of a network accessible from Internet. Later, the advent of V2X technologies provided the foundations of the so-called {\em Internet of Vehicles}, which essentially consists of a powerful sensing platform of highly sophisticated networked devices, capable to provide dedicated assistance for traffic management and pollution control through the intelligent administration and processing of environmental information \cite{Lee2016}. As a result, information from the surrounding environment around a connected car is not only limited to that from on-board sensing systems, but also includes information from other cars (e.g. inter-vehicle ranging information), road-side units, real-time traffic management systems (e.g. information concerning weather, air quality, traffic state, road conditions, parking availability, EV charging points) and even social media (e.g. tweets about events happening along a city). 

In the next future, the level of connectivity of vehicles will further increase. Enabled by the market deployment of Autonomous Vehicles (AVs), we will soon see the onset of the so-called Internet of AVs: a distributed transport network, which is based on the use of the {\em Vehicle Cloud} \cite{Lee2016}. The Vehicular Cloud is expected to allow a coordinated deployment of highly dedicated tasks such as sensor aggregation and database sharing, which is essential for the autonomous decision making process in AVs. Given its importance for AVs, there is no surprise that the Vehicle Cloud is recently stimulating much research attention. For example, filtering techniques through different communication protocols (including HTTP, TCP/IP, SMTP, WAP and Next Generation Telematics Protocol) have been proposed in \cite{He2014}, while different authors have been focusing on vehicle cloud computing. With respect to this, in e.g.  \cite{He2014} (see also the references therein), cloud computing and IoT are proposed as enablers for developing a vehicular cloud platform to intelligently share transportation-related information such as traffic management/control and road condition among different involved agents (e.g. drivers, safety authorities, regional transportation division). Related works also include: \cite{Jaworski_2011}, where a cloud computing based urban traffic control system is proposed; \cite{Olariu_2011,Hussain_2012}, where vehicular cloud architectures are proposed by integrating existing vehicular networks and on-board devices; \cite{Wang_2011}, where cloud techniques are applied to vehicular applications; and \cite{Qin_2012}, where the critical problem of establishing reliable routing among highly mobile vehicles is considered. Finally, in \cite{Lee2016}, an application is proposed for vehicle cloud to enable AVs to detect the existence of road accidents in a given geographic location.

The main idea of this paper, which builds upon the preliminary results obtained in \cite{7795769}, is inspired by the fact that, independently of the specific architecture used to deliver services to connected and autonomous vehicles, those are likely to be bombarded in the near future with information from a multitude of devices. In this context, it is very important that the vehicle is able to filter or parse this information, so that only the relevant information to the journey is passed to the driver (or to the on-board systems in the case of AVs). In essence, a key challenging requirement for both connected and autonomous vehicles is that they will need to become {\em cognitive} bodies, able to filter through a flood of continuously updated information, interpret this information, and to pass only the relevant information to the driver. In order to address this fundamental requirement, a {\em decision engine} is presented in this paper, which consists of both remote components and in-car algorithms. Then, in order to better illustrate the features of the parsing engine, this is used to develop a congestion management algorithm. The main contributions of this paper can then be summarized as follows:
\begin{itemize}
\item We first present a route parsing engine system, which filters the information that might be available to the vehicle in an intelligent manner. This is done by predicting the intentions of the driver (i.e. the likely route he/she is going to take) and parsing the real-time environmental/contextual information as a function of these intentions. Only the information that is relevant to the predicted route is then provided to the driver.
\item We build upon the route parsing engine to develop a traffic management system. Specifically, once a route is predicted for the driver, information is parsed in order to check if any obstruction is happening along the route. In the case where an obstruction is detected, the driver of the connected vehicle is notified and an alternative route is proposed if required.
\item The alternative route that is proposed to the driver is computed via a distributed algorithm, which has the goal of regulating the flow of vehicles along the obstructed route (so as not to exceed a given maximum capacity) and to balance the re-routed flow along a set of alternative routes (so that there are no alternative routes that are overloaded). We remark that the access to an affected road is now controlled in a stochastic token/buffer-like manner, rather than a deterministic regulation of the vehicular flow around a desired set point as in \cite{7795769}.
\item Finally, in addition to the use of comprehensive conventional numerical and SUMO-based simulations (similar in nature to those presented in \cite{7795769}) to test and validate our system, we now make use of the Hardware-in-the-Loop (HIL) platform originally presented in \cite{LargeScaleSUMO}. The HIL platform allows to embed a few real vehicles into scenarios created using the microscopic traffic simulation package SUMO \cite{SUMO_DLR}. As a result, the real car (onto which our algorithm is deployed) is able to interact with simulated vehicles and, in this way, the large scale effects of our distributed system can be effectively validated, as well as driver acceptability. Consequently, we provide to the driver of the real vehicle how it would actually {\em feel} to be part of the distributed system. Moreover, in order to test our system, we expanded the capabilities of the HIL and developed it in a way so it can now interface with external IoT objects. For example, the platform is now able to interface with a Twitter\textsuperscript \textregistered account through which we can generate events in the virtual environment (e.g. obstructions on some of the links in SUMO).
\end{itemize}

Before proceeding, we note that a very preliminary work on this topic was presented in the conference publication \cite{7795769}. The work reported in this present manuscript goes beyond \cite{7795769} in a number of aspects. Specifically: (i) in contrast to \cite{7795769}, a real prototype vehicle was used to test our system with network level testing of the proposed algorithms performed using a HIL platform and road tests; (ii) a new stochastic control congestion management algorithm is presented; and (iii) event obstructions are generated in real time using Twitter\textsuperscript \textregistered feeds. A final substantial difference in this present paper is that extensive simulations and road test results are presented to validate individual components of the decision engine / congestion management system.

The rest of the paper is organized as follows. In Section \ref{sec:setting}, we introduce the reference setting and goals of the proposed system. Section \ref{sec:overview}
provides a general overview of our proposed system, while Section \ref{sec:modules} provides a detailed description of each of the system modules. The experimental validation is
provided in Section \ref{sec:validation}, and the final concluding remarks are offered in Section \ref{sec:conclusions}.

\section{The Reference Setting and Goals}\label{sec:setting}

Consider the scenario where a set of $V >1$ vehicles is traveling within a given geographic area. Each of the vehicles is currently traveling along some route to reach its destination. 

At some point, one road becomes partially obstructed (or, simply, obstructed in what follows). In the context of this paper, an obstruction is an event that affects (i.e. decreases) the capacity of a given road. Obstructions might be due to unforeseen events, such as accidents, or to scheduled events, such as road maintenance. The obstruction (or event in what follows) impacts the vehicles whose route would pass through the obstructed road. As a result, the driver of those vehicles would experience a disruption to his/her trip.

\begin{figure}[h]
\centering
\includegraphics[clip, trim=6cm 2.5cm 2cm 3cm, width=3.5in]{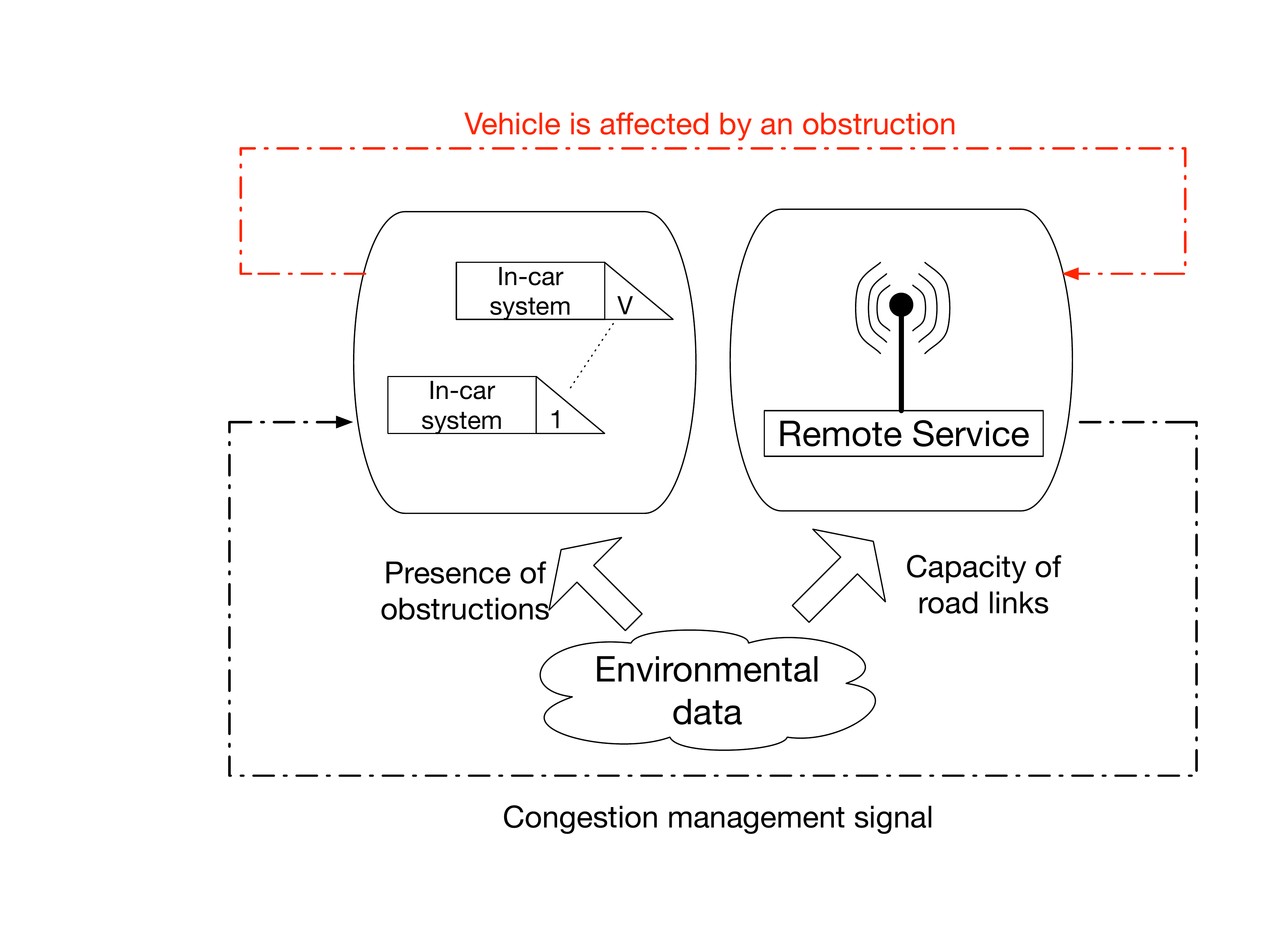}
\caption{The collaborative service for a set of $V > 1$ connected vehicles proposed in this paper. The in-car system is able to analyze environmental and contextual information in order to determine whether an obstruction will impact the the driver's trip. If this is the case, then this information is sent to a remote service which will manage the allocation of the limited resource through a congestion management algorithm based on feedback loop principles and vehicular load balancing.}
\label{Figure_scenario}
\end{figure}

The overall goal of this paper is to address the ubiquitous scenario discussed above. In particular, as outlined in Figure \ref{Figure_scenario}, we seek to design here a collaborative service between connected vehicles. The  service, which is described from a system-level viewpoint in Section \ref{sec:overview}, consists of an in-car system and of a remote service. Essentially, the in-car system for the $i$-th vehicle can sense environmental information (i.e. the presence of obstructions) and is able to estimate the most likely route that will be followed by the driver. Based on this, the in-car system parses the environmental information to check whether an obstruction will impact the trip for the driver. If this is the case, then a notification is sent to a remote service. The remote service is able to monitor the capacity of the road links within the geographic area of interest, and, based on this, implements a congestion management algorithm making use of a feedback loop mechanism, which: (i) provides to all vehicles an opportunity to access their preferred route; (ii) balances the vehicles that need to be re-routed across adjacent routes; and (iii) regulates the flow of vehicles along an obstructed road so that the capacity of that road is not exceeded. Overall, the goals of the proposed system are:
\begin{enumerate}
\item predict the next directions that the driver will take during his/her trip;
\item identify whether an obstruction has occurred along the destination that has been predicted;
\item if an obstruction is identified along the predicted route, regulate the access of the incoming vehicles to the road so that the road capacity does not exceed a given reference value;
\item in doing so, suggest and alternative route to the incoming vehicles in order to balance traffic along such routes.
\end{enumerate}

Finally, we remark that we consider two classes of road obstructions. Namely:
\begin{itemize}
\item {\em Irregular obstructions}, which typically do not happen at the same location/time and thus can not be predicted. Examples of these are lane closures due to road works or accidents. 
\item {\em Regular obstructions}, which are periodic/systematic and thus can be predicted, and resources can be regularly adjusted in advance for the duration of the obstruction. Examples of these are the closure of a school at opening/closing times or an event at an entertainment venue.
\end{itemize}

\subsection{Notation}

We now introduce the notation that will be used in the rest of the paper. In the context of this paper, the road network onto which the connected vehicles travel is characterized by a weighted directed graph, $\mathcal{G}_T$, and the associated adjacency (or connectivity) matrix, $G_T$ \cite{godsil2013algebraic}. The nodes of this graph are specific road locations such as intersections and waypoints. The links of the graph physically represent roads connecting the nodes. A link from node $l$ to node $m$ of $\mathcal{G}_T$ is denoted by $(l,m)$. Also, we denote by $\mathcal{N}^{in}_{\left(l,m\right)}$ the set of edges of $\mathcal{G}_T$ which point to edge $\left(l,m\right)$ (i.e. pointing to node $l$) and by $\mathcal{N}^{out}_{\left(l,m\right)}$ the set of edges starting from edge $(l,m)$ (i.e. departing from node $m$). The occupancy  limit associated to the link $(l,m)$ of $\mathcal{G}_T$ is denoted by $c_{(l,m)}$ and it represents the maximum capacity for that link, i.e. the maximum number of vehicles that can use that link simultaneously.

Each vehicle is characterized by its history of past trips. This historical information associated to the $i$-th connected vehicle is formalized in terms of a weighted directed graph, $\mathcal{G}^{(i)}$, and the associated connectivity matrices, $G^{(i)}$. The nodes and the edges of $\mathcal{G}^{(i)}$ are a subset of the nodes and edges of $\mathcal{G}_T$. In particular, a node/edge of $\mathcal{G}_T$ belongs to $\mathcal{G}^{(i)}$ if that node/edge has been used in the past by the $i$-th vehicle. The weight associated to $(l,m)$ is denoted by $w_{(l,m)}^{(i)}$ and is defined as the number of past trips for vehicle $i$ that included link $(l,m)$. 

An obstruction is characterized by the road where the event happened and by the corresponding maximum capacity for that road. That is, an obstruction is characterized by the tuple $\left[(l,m), c_{(l,m)}\right]$, where $(l,m)$ is the link (i.e. the road) where the obstruction happened and $c_{(l,m)}$ is the reduced capacity of that road due to the obstruction.

Finally, let $k \in \mathbb{Z}$, we denote by $x_{(l,m)}(k)$ the number of vehicles on link $(l,m)$ at time $k$. We note that the variable $x_{(l,m)}(k)$ is associated to the current state of the road network. If an obstruction occurs within the road network, then we also associate a state variable to the $i$-th vehicle. Specifically,  $y_{i}(k)$ is a binary variable that takes value $1$ if the $i$-th vehicle is granted access to the obstructed link at time $k$ and $0$ otherwise. We also denote by $\bar y_i(k)$ the average access of the $i$-th vehicle to the obstructed road up to time $k$. That is, $\bar y_i(k) := \frac{1}{k+1}\sum_{j=0}^ky_i(j)$.

\section{System Overview}\label{sec:overview}

\begin{figure}[h]
\centering
\includegraphics[clip, trim=0.5cm 2.8cm 2cm 0.8cm, width=3.5in]{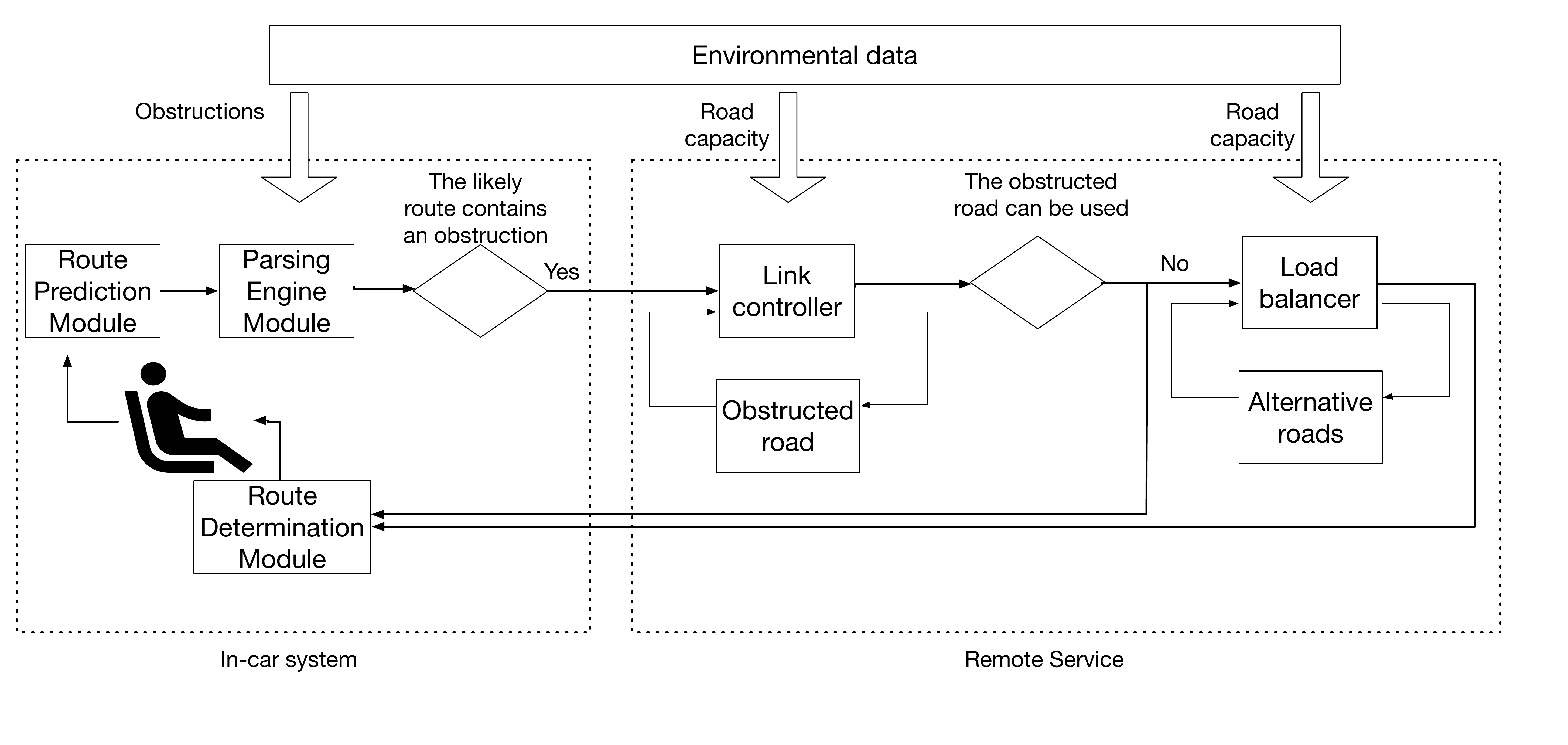}
\caption{Architecture of the system presented in this paper. In the figure, both the system modules and their functional relations are shown.}
\label{Figure_GenericArquitecture}
\end{figure}

The goal of this Section is to provide a system level view of the decision engine presented in this paper. In Figure \ref{Figure_GenericArquitecture}, the general architecture is outlined, which shows the key modules composing the system, together with their functional/logical relations. The key modules composing the architecture of Figure \ref{Figure_GenericArquitecture} and detailed in Section \ref{sec:modules} are:
\begin{itemize}
\item The \textit{Route Prediction Module}, which monitors the behavior of the driver and has the goal of predicting the most likely route that will be taken by the vehicle. The output of this module, which is part of the in-car system, is a sequence of road links which will most likely be taken by the vehicle. The prediction is performed by taking into account the history of the vehicle (i.e. historical data from past routes) and the current position of the vehicle.
\item The \textit{Route Parsing Engine}, which, based on the route predicted by the Route Prediction Module, has the goal of parsing the environmental information available to the vehicle to determine whether an obstruction can impact the trip for the driver. In the case where an obstruction happens along the route that has been predicted by the Route Prediction Module, the Route Parsing Engine outputs a notification to the Link Controller, which is hosted on the remote service.
\item The \textit{Link Controller} is part of the remote service. Such a module is able to sense environmental information (i.e. the capacity of the road links within the geographic area of interest) and receives as input the notification by the in-car system on whether a vehicle will likely pass through an obstructed link. The goal of the Link Controller is indeed that of regulating, via a feedback loop, the number of vehicles passing through a given road affected by an obstruction in order to ensure that the maximum capacity for that link is not exceeded. In particular, for any vehicle that will likely pass through the obstructed road, the controller outputs a binary variable which is equal to $1$ if the vehicle is granted access to the obstructed road, or equal to $0$ if the access is denied. In this latter case, the Load Balancer will intervene.
\item The \textit{Load Balancer} is responsible to re-route the vehicles that are not granted access to the obstructed route by the Link Controller. The Load Balancer implements a feedback loop which takes as input the capacity of the roads alternative to the obstructed link. Based on this, the Load Balancer attempts to re-route the vehicles so that the alternative links do not become over-congested, i.e. their capacities are not exceeded. The Load Balancer module returns a congestion management signal to the Route Determination Module which is part of the in-car system.
\item The \textit{Route Determination Module}, which takes as input the congestion management signal and, based on this, determines which route the vehicle should follow. This is finally notified to the driver.
\end{itemize} 

We finally remark here that the architecture of Figure \ref{Figure_GenericArquitecture} consists of a local component residing inside the connected vehicle (i.e. the in-car system) and of a global component which has access to data regarding the geographic area of interest and which might be implemented via cloud technologies (i.e. the remote service). This design choice for the architecture reflects the fact that, from the conceptual viewpoint, the goals itemized in Section \ref{sec:setting} are related to both a local and a global level. In fact, at the local level, the goal of the system is to predict the destination for the driver, detect whether an obstruction occurred along the predicted route, and suggest re-routing. The alternative routes, however, are calculated so that, at a global level, the capacity of the road affected by the obstruction does not exceed a reference value and, at the same time, traffic is balanced along the alternative routes. 

\section{The System Modules}\label{sec:modules}

In this section, we now provide a detailed description for each of the key modules outlined in Section \ref{sec:overview}.

\subsection{Route Prediction Module}\label{sec:route_prediction}

The goal of the Route Prediction Module is to estimate the next directions that will most likely be taken by the driver. That is, the module attempts to predict the route that the vehicle will take in the near future. This route is then used by the Route Parsing Engine described in Section \ref{sec:parsing} to detect whether an obstruction is going to affect the trip.
This module gathers information from past trips of the host vehicle to create a picture of where the user typically goes along a journey, and by integrating information from the current trip the prediction is further refined. 

From the functional viewpoint, the route prediction module can be broken down into three sub-modules: 1) the \textit{Edge Ranking Routine}, 2) the \textit{Edge Weighting Routine}, and 3) the \textit{Prediction Routine}.

\subsubsection{Edge Ranking Routine}

The goal of this routine is to characterize the ``importance'' of a given route from a topological viewpoint. This is done by ranking the roads (and hence the directions) taken by a driver during a journey.
Thus, the \textit{Edge Ranking Routine} takes as input the topology of $\mathcal{G}_T$ and applies a modified version of the PageRank algorithm to rank its edges, rather than its nodes \cite{Lan_Mey_12}. Specifically, let $r_{(l,m)}$ denote the ranking of the edge $\left(l,m\right)$. Then, $r_{\left(l,m\right)}$ is defined by
\begin{equation}\label{eqn:ranking}
r_{\left(l,m\right)} := \left(\left(1-d\right)+d\sum_{\left(p,q\right) \in N^{in}_{\left(l,m\right)}} r_{\left(p,q\right)}\right),
\end{equation}
where $d$ is a damping factor which, for web applications, was originally set to $0.85$ \cite{Bri_Pag_98}. In our study, inspired by \cite{Barabasi}, we set this factor to $0.93$.

\subsubsection{Edge Weighting Routine}

This routine is in charge of computing the frequency of use of a given link for the $i$-th vehicle. The input to this routine is the set of weights associated to $\mathcal{G}^{(i)}$ (recall that such a graph is constructed starting from the travel history of the $i$-th vehicle). The frequency of use of the edge $(l,m)$ of $\mathcal{G}^{(i)}$ is denoted by $f_{\left(l,m\right)}^{(i)}$ and is computed as
\begin{equation} \label{eq:Edge_Weighting}
	f_{\left(l,m\right)}^{(i)} := \frac{w_{(l,m)}^{(i)}}{\sum_{(p,q)\in\mathcal{N}^{in}_{(l,m)}}w_{(p,q)}^{(i)}}.
\end{equation}
Note that, physically, $f_{\left(l,m\right)}^{(i)}$ is the ratio between the number of times the $i$-th vehicle used the link $(l,m)$ in the past and the number of past trips that passed through node $l$.

\subsubsection{Route Prediction Routine}

Finally, this routine takes the edge rankings, $r$'s, and the frequencies, $f^{(i)}$'s, as inputs, together with the current position of the car. The output is a sequence of links that are most likely going to be visited by the vehicle. Let $\left(l,m\right)$ be the link (i.e. the road segment) onto which the car is currently traveling. We define $tr^{(i)}\left(l,W\right)$ as the {\em tree} of width $W$ (i.e. the length of the prediction horizon) generated from the node $l$ of the graph $\mathcal{G}^{(i)}$ and we denote by $br^{(i)}\left(j\right)$ the $j$-th branch of such a tree.
A score is then associated to each of the branches of $tr^{(i)}\left(l,W\right)$. Namely, the score associated to the $j$-th branch of $tr^{(i)}\left(l,W\right)$ for the $i$-th vehicle is given by
		\begin{equation} \label{eq:RoutePredict}
			R_{j}^{(i)} := \sum_{\left(m,n\right)\in br^{(i)}\left(j\right)}f_{\left(m,n\right)}^{(i)}r_{\left(m,n\right)}.
		\end{equation}

The predicted route for the $i$-th vehicle is then the one that corresponds to the branch having the maximum $R_j^{(i)}$. Note that the prediction horizon $W$ is a design parameter which affects the computational complexity of the route prediction algorithm. In the validations of this paper illustrated in Section \ref{sec:validation}, we set $W = 5$. The choice of such a value offered a good compromise between computational efficiency and prediction accuracy. The macro steps for route prediction are given in Algorithm \ref{Alg:Route_Prediction}.\\

{\bf Remark:} Note that the use of (\ref{eqn:ranking}) with $d\neq1$ is chosen
for reasons of numerical robustness and to guarantee that all the
involved edges $(p,q)$ have a ranking score $r_{(p,q)}$ different
than zero so that sink edges are avoided. Note that 
the associated transition matrix is not used for
the predictions of each vehicle $i$ but rather the connectivity matrices $G^{(i)}$, and thus every
analyzed branch (and in turn the final prediction) is always feasible.

\begin{algorithm}
                    	\caption{Route Prediction Algorithm}
                    	\label{Alg:Route_Prediction}
                    	\begin{algorithmic}[1]
				\STATE $\left(l,m\right) \gets$ current link on which the vehicle $i$ is traveling
				\STATE $tr^{(i)}\left(l,W\right) \gets$ tree of width $W$ rooted from node $l$
				\FOR{each edge $\left(p,q\right)$ in $tr^{(i)}\left(l,W\right)$}
					\STATE Get $r_{\left(p,q\right)}$ using (\ref{eqn:ranking})
					\STATE Get $f_{\left(p,q\right)}^{(i)}$ using (\ref{eq:Edge_Weighting})
				\ENDFOR
				\FOR{each branch $br^{(i)}\left(j\right)$ in $tr^{(i)}\left(l,W\right)$}
					\STATE Get $R_j^{(i)}$ using (\ref{eq:RoutePredict})
				\ENDFOR				
	\RETURN $S^{(i)} \gets$ sequence of $W$ links with largest $R^{(i)}_j$
     	        		\end{algorithmic}
\end{algorithm}

\subsection{Route Parsing Engine}\label{sec:parsing}

The objective of the \textit{Route Parsing Engine} is to parse information along the sequence of links, $S^{(i)}$, predicted by the Route Prediction Module of Section \ref{sec:route_prediction} (see Algorithm \ref{Alg:Route_Prediction}). Specifically, in the context of this paper, the goal of this module is to check, in case an obstruction occurs, if the link affected by the obstruction belongs to the sequence $S^{(i)}$.
Recall that an obstruction is characterized by the tuple $[(l,m),c_{(l,m)}]$. This module then checks whether the obstructed link $(l,m)$ belongs to $S^{(i)}$.
If this is the case, then a notification is sent to the {\em Link Controller} of Section \ref{sec:network_controller}, which, in turn, will regulate the flow of vehicles passing through the obstructed link $(l,m)$. The key conceptual steps for this module are reported in Algorithm \ref{Alg:Route_Parsing}. Note that this algorithm returns a binary variable $C^{(i)}$ which can take value $0$ if the obstruction does not belong to the sequence $S^{(i)}$ or it can take value $1$ if the obstruction belongs to $S^{(i)}$. The variable $C^{(i)}$ is taken as input by the Link Controller. In particular, if $C^{(i)} = 1$, then the Link Controller is activated.

\begin{algorithm}
                    	\caption{Route Parsing Engine Algorithm}
                    	\label{Alg:Route_Parsing}
                    	\begin{algorithmic}[1]
				\STATE Get $S^{(i)}$ from \textit{Route Prediction Module} (Algorithm \ref{Alg:Route_Prediction}) 
				\STATE Gather environmental data
				\STATE $C^{(i)} \gets 0$ 
	           			\IF{obstruction $[(l,m),c_{(l,m)}]$ is detected}
	           			\IF{$(l,m)\in S$}
					\STATE $C^{(i)}\gets 1$
						\ENDIF
						\ENDIF
						\RETURN $C^{(i)}$	
     	        		\end{algorithmic}
\end{algorithm}

\subsection{Link Controller}\label{sec:network_controller}

This module is invoked whenever an obstruction is detected, i.e. whenever the output variable, $C^{(i)}$, of the Route Parsing Engine is equal to $1$.
We assume that the link affected by the obstruction is $(l,m)$ and we build a mechanism around such a link to regulate access of vehicles to $(l,m)$,
so that $c_{(l,m)}$ is not exceeded. The output of this module is the probability for the vehicles demanding allocation to link $(l,m)$.
We propose here two Link Controllers: the first is for irregular obstructions, while the second is for regular obstructions. The use of two different controllers
for such cases is motivated by the fact these two types of obstructions are  fundamentally different in nature from each other. In particular,
we can leverage the predictability of regular obstructions to design more tailored algorithms. 

\subsubsection{Allocation through Irregular Obstructions}

We start with considering the case of an irregular obstruction along the link $(l,m)$. The goal is to regulate access to this link so that $x_{(l,m)}(k) \le c_{(l,m)}$, $\forall k$.
In order to do so, the link controller returns a signal, $P_{(l,m)}(k)$, to the {\em Route Determination Module} of the in-car system of the vehicles demanding allocation.
This signal physically represents the probability for a vehicle to be allocated to $(l,m)$. Such a probability is computed as
\begin{equation} \label{eq:Controller1}
	P_{(l,m)}\left(k\right) := \begin{cases}0, & \text{if } e_{(l,m)}\left(k\right) \leq 0, \\ 
	\frac{e_{(l,m)}\left(k\right)}{c_{(l,m)}}, & \text{if } 0<e_{(l,m)}\left(k\right) < c_{(l,m)}\left(k\right),\\ 
	1, & \text{if } e_{(l,m)}\left(k\right) = c_{(l,m)}\left(k\right), \end{cases}
\end{equation}
where $e_{(l,m)}\left(k\right) = c_{(l,m)} - x_{(l,m)}\left(k\right)$. 

\subsubsection{Allocation through Regular Obstructions}

Now, we analyze the case of regular obstructions. As we describe below, the predictability of those obstructions can be leveraged to have a more tailored design, which
e.g. includes balancing, over time, the average access of drivers to the obstructed road. In particular, the closed loop mechanism we design, which offers a generalization of the one proposed in \cite{QoS_Parking}, allows to regulate access of vehicles to the obstructed route in a way such that, on average in the long run, marginal cost of each of the vehicles associated to not using the obstructed link is the same. Moreover, the marginal cost for each of the vehicles will be dependent on the available capacity along the obstructed link. In the case of regular obstructions, the link controller calculates a time-varying coefficient, $\gamma_{(l,m)}$, which is broadcast to the all in-car system of the vehicles that are likely going to pass through the obstructed link $(l,m)$. The coefficient $\gamma_{(l,m)}$ is calculated as follows:

\begin{equation}\label{Eq:GammaEvolution}
	\gamma_{(l,m)}\left(k+1\right) = \alpha \gamma_{(l,m)}\left(k\right) + (1-\alpha)g\left(e_{(l,m)}\left(k\right)\right),
\end{equation}
where, $0<\alpha<1$ is a design parameter  and the function $g: \mathbb{R} \rightarrow [0,1]$ is a piece-wise continuous function defined as
\begin{equation}\label{Eq:FuntionGModified}
	g(e_{(l,m)}) :=  \begin{cases}0, & \mbox{if  } e_{(l,m)} \leq 0,\\
	\frac{e_{(l,m)}}{c_{(l,m)}}, & \mbox{if  } 0 < e_{(l,m)} < c_{(l,m)},\\
	1, & \mbox{if  } e_{(l,m)} = c_{(l,m)}. \end{cases}
\end{equation}

Clearly, the dynamical system in (\ref{Eq:GammaEvolution}) is a stable discrete-time dynamical system and it converges towards the value $\bar g = g(e_{(l,m)})$. 

The macro steps for the Link Controller are given in Algorithm \ref{alg:link_controller}.

\begin{algorithm}
                    	\caption{Link Controller Algorithm}
                    	\label{alg:link_controller}
                    	\begin{algorithmic}[1]
                    	\IF{$C^{(i)} \mbox{ is } 1$}
                    	\STATE $c_{(l,m)} \gets$ capacity of obstructed link
                    	\STATE $x_{(l,m)}(k) \gets$ number of cars on the obstructed link
                    	\IF{Irregular obstruction}
                    	\STATE Get $P_{(l,m)}(k)$ using (\ref{eq:Controller1})
                    	\RETURN $P_{(l,m)}(k)$
                    	\ELSE
                    	\STATE Get $\gamma_{(l,m)}\left(k\right)$ using (\ref{Eq:GammaEvolution})
                    		\RETURN $\gamma_{(l,m)}\left(k\right)$
                    	                       	\ENDIF
                    	\ENDIF
     	        		\end{algorithmic}
\end{algorithm}

\subsection{Load Balancer}

The load balancer is responsible for calculating a set of probabilities associated to links which are alternative to the obstructed link $(l,m)$. We assume that, for each link of $\mathcal{G}_T$, this module has built-in a set of $N_{(l,m)}$ alternative links. Let $\mathcal{A}_{(l,m)} = \left\{(l,m)_1,\ldots,(l,m)_{N_{(l,m)}}\right\}$ be the set of links alternative to link $(l,m)$. We denote the cardinality of $\mathcal{A}_{(l,m)}$ by $\vert \mathcal{A}_{(l,m)}\vert$, i.e. $\vert \mathcal{A}_{(l,m)}\vert = N_{(l,m)}$.  We build a feedback loop which, based on the state of the links of $\mathcal{G}_T$, allocates the vehicles incoming to but not allocated link $(l,m)$ to one of the possible alternatives in $(l,m)_j$, $j=1,\ldots,N_{(l,m)}$. Specifically, the output of this module is a set of time-varying probabilities, say $\left\{P_{(l,m)_{j}}\left(k\right)\right\}_{j=1,...,N_{(l,m)}}$, associated to each of the alternative links in $\mathcal{A}_{(l,m)}$.

We start with considering the case when the number of vehicles along the alternative link $(l,m)_j$ is different from $0$, $\forall j=1,\ldots,N_{(l,m)}$.
That is, we start with considering the case where $x_{(l,m)_j}\left(k\right) > 0$, $\forall j= 1,\ldots,N_{(l,m)}$. In this case,
the probabilities associated by the Load Balancing module for the possible $N$ alternatives to the obstruction at the link $(l,m)$ are given by
\begin{equation}
\label{eq:Re-Routinga}
\begin{array}{*{20}l}
P_{(l,m)_j}(k) = \frac{1/h_{j}(k)}{\sum_{q=0}^{N_{(l,m)}}\left(1/h_{q}(k)\right)}, \\
h_{j} (k) = \frac{x_{(l,m)_j}\left(k\right)}{\sum_{q=0}^{N_{(l,m)}}x_{(l,m)_q}\left(k\right)},
\end{array}
\end{equation}
where, physically, $h_{j}(k)$ can be seen as the time-varying portion of the total load of vehicles which is passing through the alternative links $(l,m)_j$.
Note that, in essence, the vector $P_{(l,m)_j}(k)$ is the same for all the vehicles forced to reroute at instant $k$, and it only depends on the current state of link $(l,m)_j$.

Instead, in the case when there are some alternative links having no vehicles, i.e. when there exists some $(l,m)_j \in \mathcal{A}_{(l,m)}$ such that $x_{(l,m)_j}\left(k\right)=0$,
then the probabilities are computed as
\begin{equation}
\label{Equ:LBEG3}
P_{(l,m)_j}(k)=\left\{ \begin{array}{cc}
\frac{1} {N^{(0)}_{(l,m)}(k)}, & \forall (l,m)_j: \ x_{(l,m)_j}\left(k\right) = 0,\\
0, & \text{ otherwise},
\end{array}\right.
\end{equation}
where $N^{(0)}_{(l,m)}(k)$ is the number of links alternative to $(l,m)$ having no vehicles at time $k$. That is, $N^{(0)}_{(l,m)}(k) := \vert (l,m)_j \in \mathcal{A}_{(l,m)}: x_{(l,m)_j}(k) = 0 \vert$. 

The value $P_{(l,m)_j}$ obtained either via (\ref{eq:Re-Routinga}) or (\ref{Equ:LBEG3}) is finally broadcast to the $i$-th vehicle, for which the route predicted by the Route Prediction Module is affected by the obstruction. Specifically, such values are received by the {\em Route Determination Module} described below. The macro steps for the Link Controller are given in Algorithm \ref{alg:load_balancer}.

\begin{algorithm}
                    	\caption{Load Balancer Algorithm}
                    	\label{alg:load_balancer}
                    	\begin{algorithmic}[1]
						\STATE $(l,m) \gets $ obstructed road link
						\STATE $\left\{(l,m)_1,\ldots,(l,m)_{N_{(l,m)}}\right\} \gets$ $N_{(l,m)}$ alternative links  to $(l,m)$                    	                   	
						\FOR{$j\in1,\ldots,N_{(l,m)}$}
						\STATE Get $x_{(l,m)_j}(k)$
						\IF{$x_{(l,m)_j}(k) >0$ for all alternative links}
						\STATE Get $P_{(l,m)_j}(k)$ using (\ref{eq:Re-Routinga})
						\ELSE
						\STATE Get $P_{(l,m)_j}(k)$ using (\ref{Equ:LBEG3})
						\ENDIF
						\ENDFOR
						\RETURN vector $\left\{P_{(l,m)_{j}}\left(k\right)\right\}^{j=1}_{N_{(l,m)}}$
     	        		\end{algorithmic}
\end{algorithm}

\subsection{Route Determination Module}

The Route Determination Module is responsible to determine, given the outputs of the Link Controller and of the Load Balancer,
which route will be taken by the vehicle that is predicted to be affected by the obstruction along the link $(l,m)$. For the sake of clarity,
we separately discuss the case of irregular obstructions and the case of regular obstructions.

\subsubsection{Route determination for an irregular obstruction}

In this case, the Route Determination Module of the $i$-th vehicle affected by the obstruction receives as input the $[N_{(l,m)}+1]$-dimensional vector of probabilities computed by the Link Controller and by the Load Balancer
\begin{equation}\label{eqn:prob_vector}
\left[ P_{(l,m)}(k) , \left\{P_{(l,m)_{j}}\left(k\right)\right\}^{j=1}_{N_{(l,m)}} \right].
\end{equation}

Essentially, the road to which the $i$-th vehicle is allocated is determined by flipping a coin against the probabilities in (\ref{eqn:prob_vector}). Specifically:
\begin{itemize}
\item First, a coin is flipped against the probability $P^{(i)}_{allocation}=P_{(l,m)}$. If this coin toss is successful, then the $i$-th vehicle is allocated to the obstructed link $(l,m)$.
\item If the coin toss is not successful, then the $i$-th vehicle is not granted access to the link $(l,m)$ and the vehicle is allocated to one of the alternative links. In order to allocate the vehicle to one alternative link, a coin is again flipped, this time against the probabilities $\left\{P_{(l,m)_{j}}\left(k\right)\right\}^{j=1}_{N_{(l,m)}}$.
\end{itemize}

\subsubsection{Route determination for a regular obstruction}

In case of a regular obstruction, the Route Determination Module receives as input the $[N_{(l,m)}+1]$-dimensional vector
\begin{equation}\label{eqn:prob_vector_2}
\left[ \gamma_{(l,m)}(k) , \left\{P_{(l,m)_{j}}\left(k\right)\right\}^{j=1}_{N_{(l,m)}} \right].
\end{equation}

The first step is then to devise the probability $P^{(i)}_{allocation}$ associated to the $i$-th vehicle to be allocated the link affected by the regular obstruction (as
 $\gamma_{(l,m)}$ is not indeed a probability). Thus, in order to do so, and inspired by \cite{QoS_Parking}, we let:
\begin{equation}\label{eq:FairP}
P^{(i)}_{allocation} = \gamma_{(l,m)}(k)H_i(k).
\end{equation}

In the above equation, the dynamics $H_i(k)$ is specific to the $i$-th vehicle and is given by
\begin{equation}\label{eq:HistFunc}
H_i\left(k\right) = 
\frac{\bar{y}_{i} \left(k\right)}
 {\phi_{i}\left({\bar{y}_{i}} \left(k\right) \right)},
\end{equation}
where (i) $\bar{y}_{i}$ is the average value of allocation of resources for user $i$, $y_{i}$, and (ii) $\phi_{i}\left(\cdot\right)$ is an strictly increasing function (on the domain of interest).
Once $P_{allocation}^{(i)}$ is obtained from (\ref{eq:FairP}), then the coin toss mechanism described in the case of irregular obstructions is applied. The macro steps for the Route Determination module are given in Algorithm \ref{alg:route_det}.

\begin{algorithm}
                    	\caption{Route Determination Algorithm}
                    	\label{alg:route_det}
                    	\begin{algorithmic}[1]
						\STATE Get $\left\{P_{(l,m)_{j}}\left(k\right)\right\}^{j=1}_{N_{(l,m)}}$ from Load Balancer
						\IF{Regular Obstruction}
							\STATE Get $\gamma_{(l,m)}(k)$ from Link Controller
							\STATE Get $P^{(i)}_{allocation}$ using (\ref{eq:FairP}) and (\ref{eq:HistFunc})
						\ELSE
							\STATE Get $P_{(l,m)}(k)$ from Link Controller
							\STATE Set $P^{(i)}_{allocation}=P_{(l,m)}(k)$
						\ENDIF
						\STATE Coin Toss against $P^{(i)}_{allocation}$
						\IF{Coin Toss successful}
							\STATE Grant access of the obstructed link $(l,m)$ to vehicle $i$
						\ELSE
							\STATE Coin toss against $\left\{P_{(l,m)_{j}}\left(k\right)\right\}^{j=1}_{N_{(l,m)}}$
							\STATE Grant access of vehicle $i$ to the alternative link for which the coin toss is successful
						\ENDIF				
												\RETURN link to be used by the $i$-th vechicle
     	        		\end{algorithmic}
\end{algorithm}

\section{Experimental Validation}\label{sec:validation}

The experimental validation of the proposed methodology is divided into two categories: 1) irregular obstructions, and 2) regular obstructions.
For each of the above categories, different tools and scenarios are used. The corresponding setups and obtained results are presented in the following subsections.

\subsection{Irregular Obstructions}

Here, the evaluation process is performed through two different methods:
1) conventional traffic simulations, and 2) hardware-in-the-loop (HIL) validation. Both validation
methods are based on the use of the open-source, microscopic traffic simulator SUMO \cite{SUMO_DLR},
in collaboration with Python scripts to interact online with the simulations using the {\it Traffic Control Interface} (TraCI) package\footnote{http://sumo.dlr.de/wiki/TraCI}. 
SUMO enables for a realistic reconstruction of a given vehicular traffic problem, and allows for the implementation
of a vehicle to vehicle/infrastructure (v2x) -like solutions via TraCI commands in Python environment.

SUMO simulations are, in general, comprised of two mandatory layers, namely:
\begin{enumerate}
\item Network Layer: the simulated road infrastructure (i.e. road links, junctions, connections, traffic lights, etc);
\item Vehicular Layer: the vehicles embedded into the simulation and their associated parameters/behaviors (i.e. vehicle types, routes, emission classes, etc).
\end{enumerate}
An additional (not mandatory) layer is the Shapes Layer, which serves among others to decorate
the simulation environment when the graphical user interface is used (e.g., polygons representing
areas of interest such as buildings, and points-of-interest representing locations of interest).

\subsubsection{General setup}\label{s:GenSetup}

\begin{figure}[h]
\centering
\includegraphics[width=3.0in]{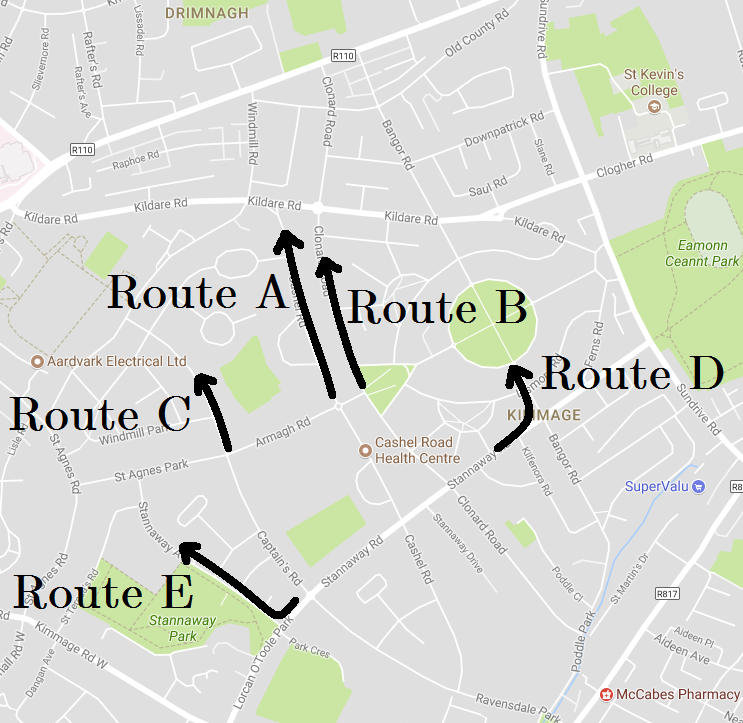}
\caption{Selected area between Drimnagh and Kimmage, Dublin, and main routes used to study the case of irregular obstructions.}
\label{Fig03SelectedArea}
\end{figure}

Concerning the Network Layer, we selected the area between Drimnagh and Kimmage, Dublin,
as depicted in Figure \ref{Fig03SelectedArea}, and imported the corresponding road network from OpenStreetMaps into SUMO
environment. For the Vehicular Layer, we defined a single vehicle type with the following parameters:
\begin{itemize}
\item Length: 5 m,
\item Acceleration: 0.8 m$^2$/s,
\item Deceleration:  4.5 m$^2$/s,
\item Car-following model: Krauss.
\end{itemize}
Additionally, we generated 481 possible routes which cross the selected area from side to side, and we created
a random number of vehicles using these routes. Finally, we define a point on interest to show the location of the obstruction,
and polygon to highlight the affected area by the obstruction, i.e. the radius around the obstruction at which the cars are
assessed for risk-mitigation. In this investigation, the aforementioned radius was set about 575 m to ensure both that the resulting
delay between the vehicles being allocated resources and vehicles clearing the obstruction is short, and to allow more alternative
routes to be taken.

Since the number of $N$ links in the imported road network determines the dimensions $N \times N$
of the associated connectivity matrix $G_T$ in question, the operations involving
$G_T$ (e.g., finding the solution of a linear system, required for the modified PageRank algorithm) will result in a high computational load
for a large $N$. In the case of the selected area, we have that $N=6126$, which will lead to a $6126 \times 6126$
dense matrix $G_T$. Due to this challenging fact, we use instead 17 different matrices $G_T$, each of them corresponding
to one of the possible 17 road links on the boundary of the selected area. This simple action allows us to reduce the algorithmic burden
as each of these matrices can be handled as a sparse matrix with much fewer entries (compared to the  $6126 \times 6126$
dense matrix $G_T$). The construction of each of these connectivity matrices was conducted by generating a random number of trips departing from the given boundary road link, and then by storing the total number of times that every involved link has been used.

\subsubsection{Obstruction generation} \label{s:ObstructionGen}

The partial obstruction is represented in SUMO environment as a road link with a reduced speed limit (concerning the configuration parameters), which is also
related to a given maximum capacity (concerning the setpoint for the Link Controller). We designed an interactive method to generate, in real time, the obstruction on any arbitrary road link. Such a method is based on the data analysis (text mining) of tweets from Twitter\textsuperscript \textregistered reporting about traffic incidents. This is possible via the Twitter\textsuperscript \textregistered Streaming API accessed by
{\it tweepy}\footnote{http://www.tweepy.org/}, a Python-based library. For this, we designed the dedicated Twitter\textsuperscript \textregistered account ``SmartTransport'' (\texttt{@smart\_tran}),
and define the format for the posted tweets to open an incident as
\[
<\mbox{T1}>+<\mbox{T2}>+<\mbox{T3}>+<\mbox{T4}>+<\mbox{T5}>
\]
where
\begin{itemize}
\item T1: ``$\mbox{New road incident:} <Location>.\,\,$'',
\item T2: ``$\mbox{LatLon:} <latitude,longitude>.\,\,$'',
\item T3: ``$\mbox{Maxcapacity:} <integer>.\,\,$'',
\item T4: ``$\mbox{Maxspeed:} <speed>\,\,\mbox{[km/h]}.\,\,$'',
\item T5: ``$\mbox{Time:} <timestamp>.\,\,$''.
\end{itemize}

Similarly, we define the format for the posted tweets to close an incident as
\[
<\mbox{T6}>+<\mbox{T7}>
\]
where
\begin{itemize}
\item T6: ``$\mbox{Road incident closed:} <Location>.\,\,$'',
\item T7: ``$\mbox{Time:} <timestamp>.\,\,$''.
\end{itemize}

With this method, the nearest road link in SUMO will be selected as the affected link once a new road incident is posted, and its maximum speed
will be set as the provided ``Maxcapacity'' until such an incident in closed. In our tests, the location of the incident was selected as
Cashel Rd North (latitude: 53.322340, longitude: -6.306612), the maximum capacity in $\left\{3,4,5,6\right\}$, and the speed limit at 1.5 km/h.

\subsubsection{Results from numerical simulations}

\begin{figure}[h]
\centering
\includegraphics[clip, trim=1cm 1.5cm 1.4cm 1cm, width=3.5in]{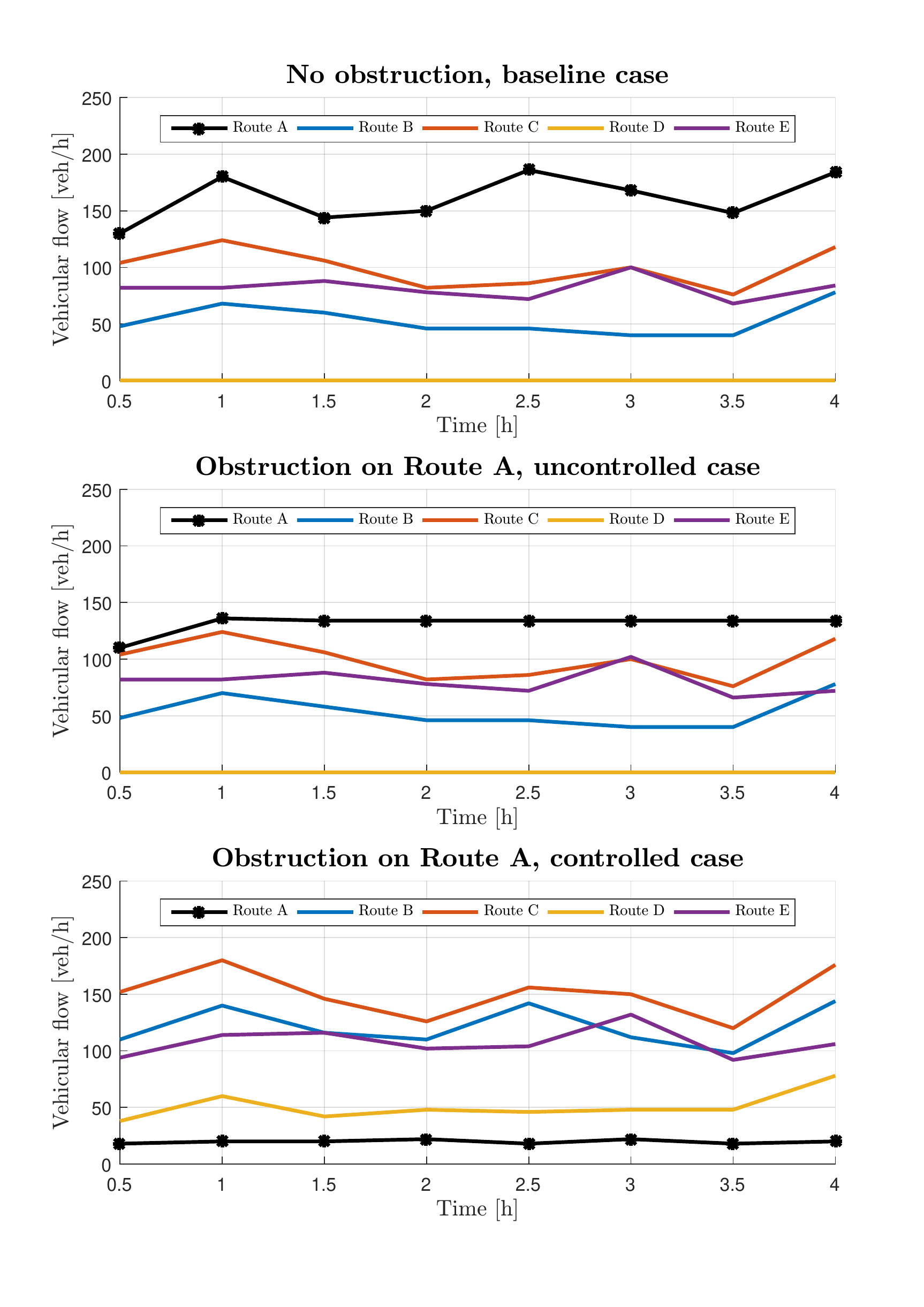}
\caption{Irregular obstruction: Vehicular flows observed for the selected routes. [Setup: incident on Cashel Rd North (included in Route A), maximum capacity equal to 3, speed limit at 2 km/h.]}
\label{Fig04IrregularFlows}
\end{figure}

\begin{figure}[h]
\centering
\includegraphics[clip, trim=1cm 1.8cm 1cm 1cm,width=3.5in]{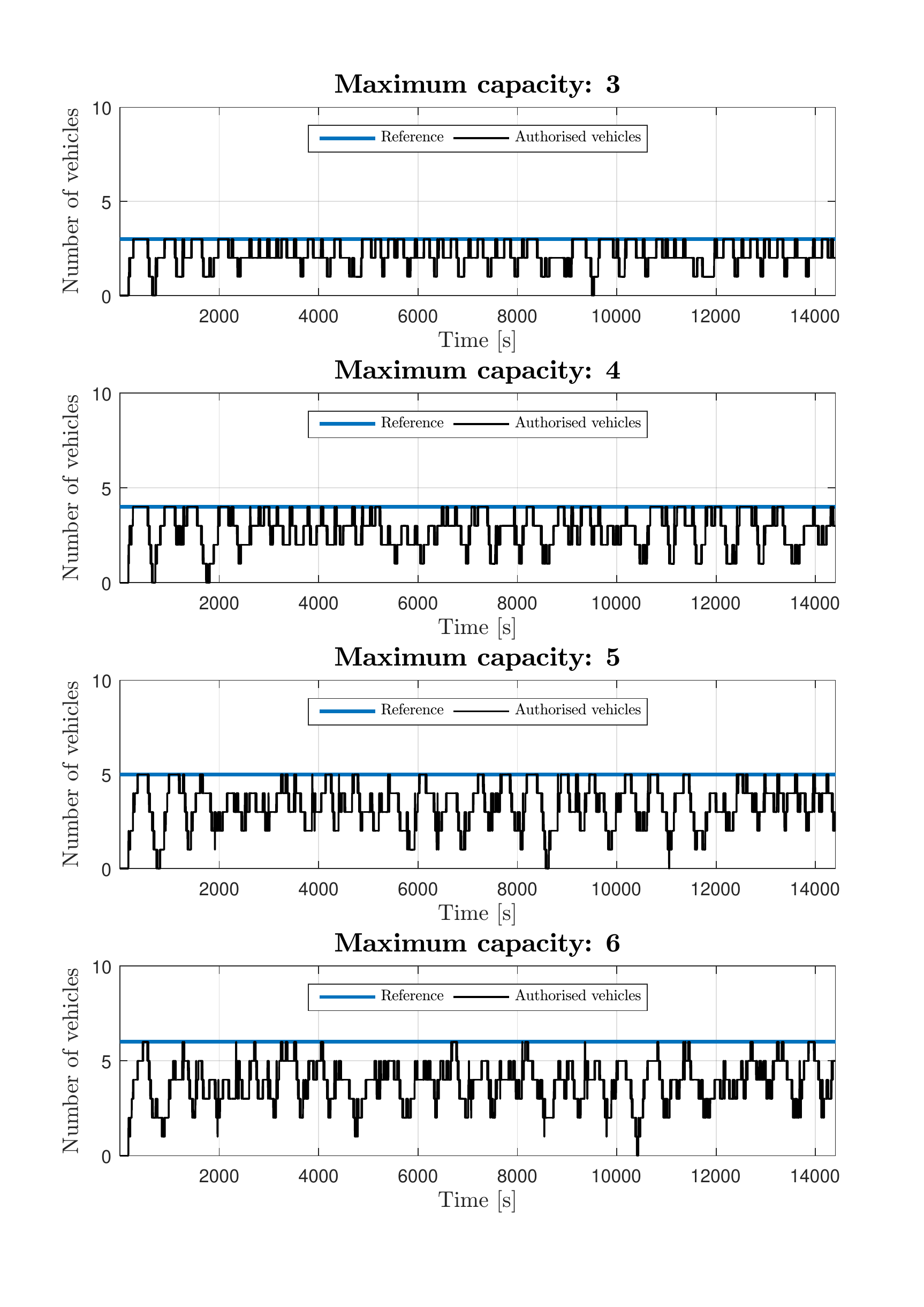}
\caption{Irregular obstruction: number of vehicles with access to the obstructed link, as the result of the Link Controller action. [Setup: incident on Cashel Rd North, maximum capacity in $\left\{3,4,5,6\right\}$.]}
\label{Fig05IrregularControl}
\end{figure}

We performed a number of simulations for 4 different scenarios corresponding to random realizations of the experiment for different
values of the maximum capacity in $\left\{3,4,5,6\right\}$. Each of these simulation corresponded to a period of 4 hours,
with the partial obstruction included from the beginning in the uncontrolled/controlled cases. The results of these experiments are presented
in Figures \ref{Fig04IrregularFlows} and \ref{Fig05IrregularControl}. Figure \ref{Fig04IrregularFlows}
shows the vehicular flow in three different scenarios: (a) no obstruction, baseline test, (b) obstruction, uncontrolled case (i.e. no link control, no load balancing),
(c) obstruction, controlled case (i.e. link control and load balancing).
As it can be seen from the baseline simulation, Route A is used by a large amount of vehicular traffic in comparison with the other routes. Thus, the vehicular flow
through it gets saturated when it becomes partially obstructed in the uncontrolled case. If the link control and the load balancer are activated at the beginning of the
obstruction, it can be seen that the vehicular traffic through Route A is reduced according to the specifications of the link controller, and the excess of the vehicular
traffic is successfully distributed to the alternative routes (i.e. the traffic flow through them became closely matched). The performance of the link controller for different
values of the maximum capacity is presented in Figure \ref{Fig05IrregularControl}, where it is clear that the controller is able to maintain the traffic less or equal than the reference value.\\

{\bf Remark:} In a real-world implementation, the in-car system requires to interact with the driver so that the output from the Route Prediction Algorithm
can be verified. In our experiments, we emulate this for all the cars by checking if the obstructed road belongs to the host vehicle's route (i.e. we check whether an element belongs to an array) which is an accurate process; due to this fact, the Route Parsing Engine will always be completely sure whether the obstruction will affect the trip, and so the Link Controller will not have any disturbances. However, if real users fail in providing a correct verification (e.g., they make a mistake, or they just lie), then the Link Controller will experience disturbances as some vehicles will not be rerouted, and thus the number of vehicles on the obstruction will eventually surpass the reference (see Fig. \ref{Fig05IrregularControl}).\\

\subsubsection{Results from HIL validation}

For a more realistic assessment of the proposed methodology, involving the user response to the traffic management system,
we propose the use of the HIL approach presented in \cite{LargeScaleSUMO}.
Since the only real-time information required from the target car is the position and heading angle, then we designed an ad-hoc web service
based on the HTML Geolocation API\footnote{https://www.w3schools.com/html/html5\_geolocation.asp}, rather than designing a smartphone app
which interacts with an OBD2 interface (as done in \cite{LargeScaleSUMO}). This web service has two main functions:
\begin{itemize}
\item automatically send (with a given fixed frequency) the GPS position and heading angle of the target car to the HIL platform; and
\item present to the diver the instructions received from the HIL platform (i.e. re-route directions).
\end{itemize}

\begin{figure}[h!]
\centering
\subcaptionbox{Access to the limited resource is allowed. \label{subfig:HIL_NoRec}}
{\includegraphics[width=3.0in]{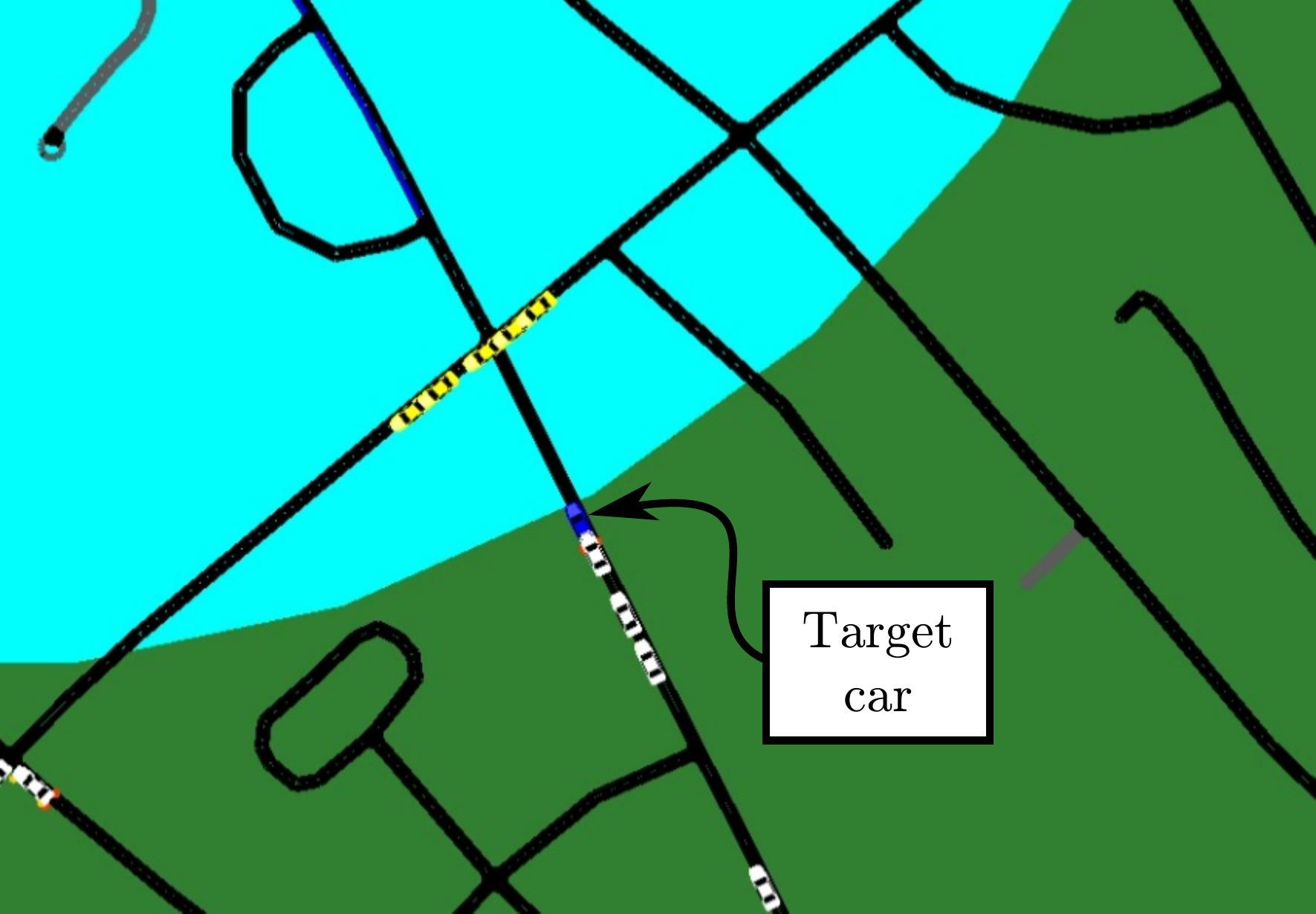}}
\subcaptionbox{Access to the limited resource is denied.\label{subfig:HIL_Rec}}
{\includegraphics[width=3.0in]{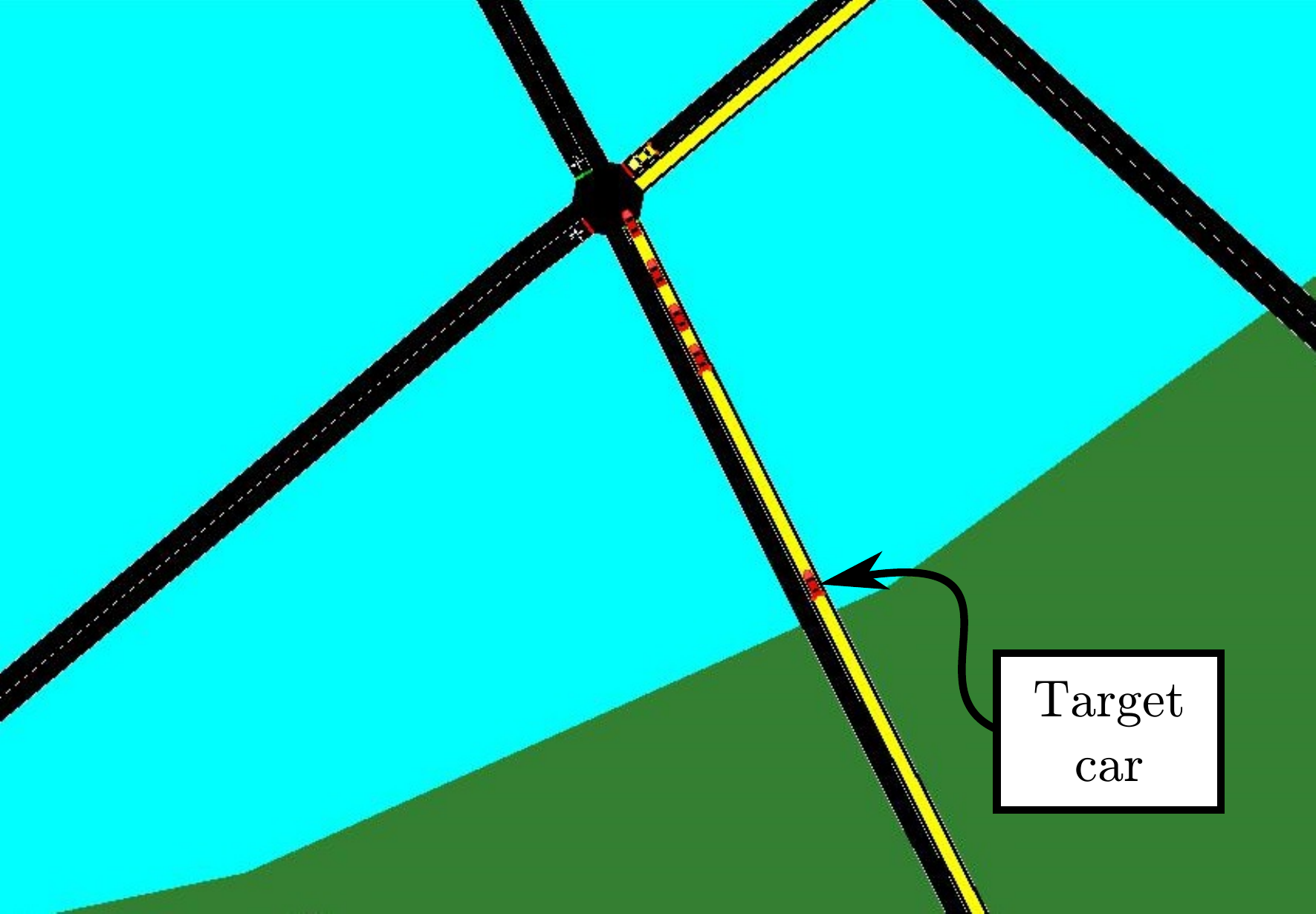}}
\caption{Results from the HIL tests: (a) the target car is allowed to use the affected road link (associated car's avatar is coloured blue), and (b) the target car must be re-routed (associated car's avatar is coloured red), and thus new driving directions (yellow line) are generated. [Light blue circle: area affected by the obstruction (see Section \ref{s:GenSetup}).]}\label{FigHILtest}
\end{figure}

The HIL validation was conducted using the general setup described in Section \ref{s:GenSetup}, and with a 2015 Toyota Prius 1.8 VVT-I as the test car. Additionally, we started a HIL simulation and few instants later created an obstruction on Cashel Road North (latitude: 53.322334, longitude: -6.306720) using our obstruction generation method (see Section \ref{s:ObstructionGen}, and \url{https://twitter.com/smart_tran} for illustration). We then embedded the Prius at a location in the south of the selected area (shown if Figure \ref{Fig03SelectedArea}), and drove the car toward the affected road link.
Different experiments were performed. In most cases, the test car was predicted to use the obstruction and allowed to use it (as a result of each particular traffic condition), as shown in Fig. \ref{subfig:HIL_NoRec}. An example of the opposite is presented in Fig. \ref{subfig:HIL_Rec}, where target car is not accepted to go through the obstruction and thus re-routing directions are generated and sent to the driver via the ad-hoc web service.

Concerning the driving experience, we were able to verify that, as it was designed and implemented, the system only intervened when it was necessary and thus we received only the relevant information, so that the interactions with the the system (and thus the distractions) were minimal.

\begin{figure}[h!]
\centering
\includegraphics[width=3.5in]{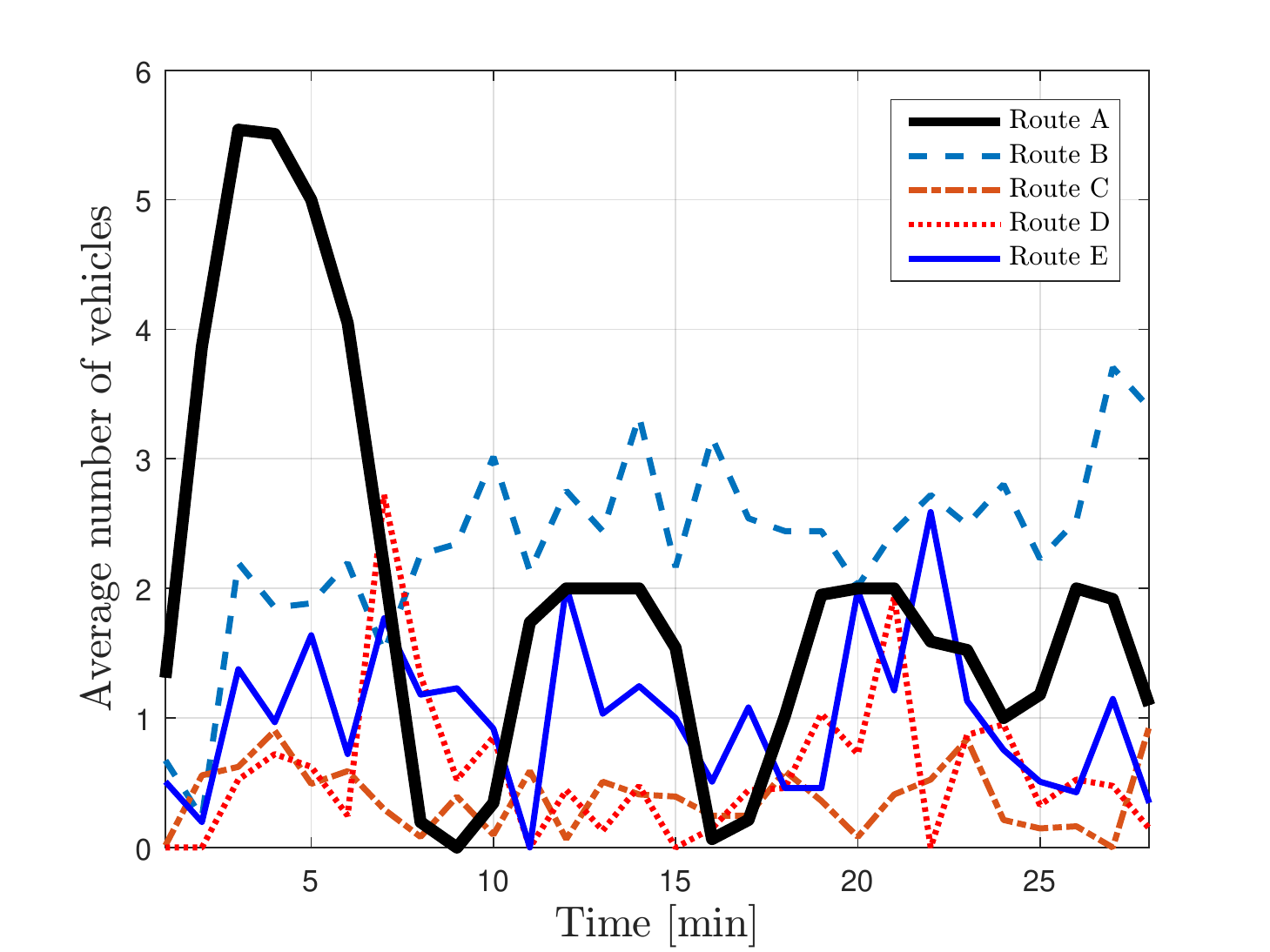}
\caption{Vehicular traffic information from HIL validation: Evolution of the average number of vehicles on the obstructed route (Route A) and the alternative routes (all the others). Obstruction was generated at around time 2 min. }
\label{Fig07TrafficInfoHIL}
\end{figure}

Finally, concerning the performance of the traffic controller and the load balancer, we can see from Fig. \ref{Fig07TrafficInfoHIL} (despite the short time span of the test) that the average number of vehicles on each involved route is consistent with the design:
\begin{itemize}
\item the route with the obstructed link (i.e. Route A) reaches a state in which it is occupied with no more than 3 vehicles (as defined in the test tweet) ;
\item the alternative routes (especially Routes B and D) account for the vehicular traffic not allowed to use Route A.
\end{itemize}

\subsection{Regular Obstructions}

Due to the regularity of these kinds of obstructions, these were studied on a longer time-span than the irregular obstructions: a period of 16 hours per day,
repeated over the course of a year (i.e. 365 times). Due to this fact, we do not use the traffic simulator SUMO but make use of a
traditional numerical simulation in Python with the following setup:
\begin{itemize}
\item We define the simulation step $k$ as 1 s, and emulate 2 access requests per 16-hour period for the participating vehicles.
\item We assume the error $e_{\left(l,m\right)}\left(k\right)$ used for the link controller as a 1-dimensional random walk between 0 and $c_{\left(l,m\right)}$,
with $c_{\left(l,m\right)}\in{3,6}$; we use a walking step given by $\left\lfloor\beta r\left(k\right)\right\rceil$ where $r\left(k\right)$ is a normally distributed random number,
$\left\lfloor\bullet\right\rceil$ is the nearest integer function, and $\beta$ is a design parameter to be adjusted. An empirical tuning process showed that
$\beta=0.35$ provides similar values of $e_{\left(l,m\right)}\left(k\right)$ as those obtained in the case of irregular obstructions.
\item We analyze the performance of 10 cars, and use $\alpha=0.1$ for Eq. \ref{Eq:GammaEvolution} and $\phi_{i}\left(z\right)=4z^3$ for the calculation of Eq. \ref{eq:HistFunc}.
\end{itemize}

In the experiments, when a user $i$ is required to compete for the affected link, their allocation history $y_i$ is queried to calculate the average
allocation $\bar{y}_i$ until the current simulation step so that Eq. \ref{eq:HistFunc} can be solved, and in turn Eq. \ref{eq:FairP}.
The allocation of the resource is then evaluated through a coin toss against $P^{(i)}_{allocation}$, after which $y_i$ is updated in the following manner:
if user $i$ gets permission is allocated the resource, we append 1 to $y_i$; otherwise, we append 0.\\

{\bf Remark:} Note that re-routing in the case of allocation denial does not affect (on average) the calculation of $P^{(i)}_{allocation}$,
and thus it is not explicitly emulated in our experiments.\\

\begin{figure}[h]
\centering
\includegraphics[clip, trim=0.4cm 0.5cm 0.8cm 0.5cm, width=3.5in]{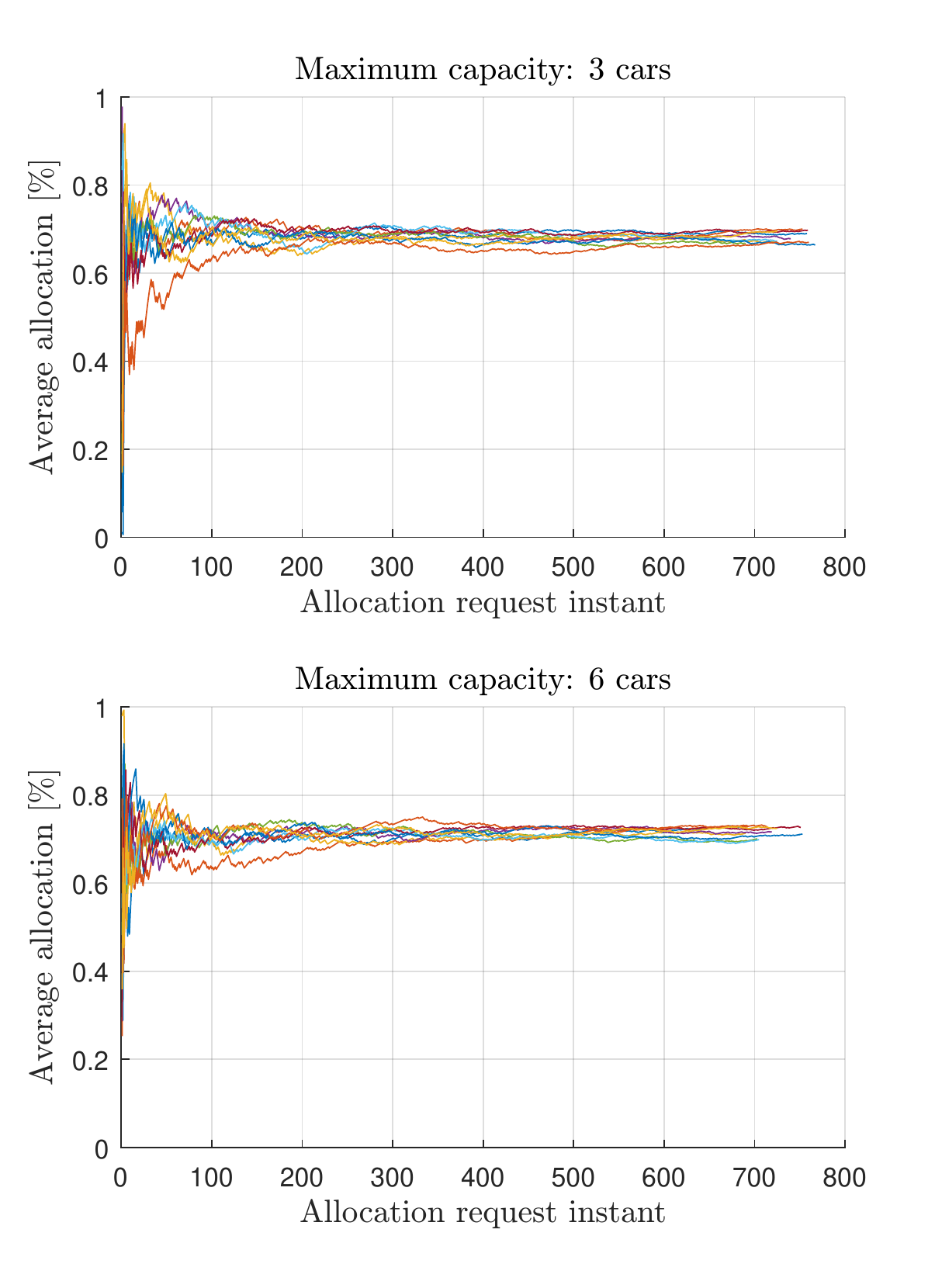}
\caption{Regular obstruction: Evolution of the average allocation (percentage) for the 10 simulated vehicles under study, using different values for the maximum capacity. }
\label{Fig08RegularAllocationNew}
\end{figure}

The numerical results from the above experiments are presented in Figure \ref{Fig08RegularAllocationNew} .There, it can be noticed that all the average allocation values of the analyzed vehicles show convergence to a similar value 
(which is slightly higher in the case of a higher maximum capacity allowed for the regular obstruction), empirically showing the fairness of the proposed approach.

\section{Conclusions and Future Work}\label{sec:conclusions}

In this paper we designed, tested and validated a decision engine system, which is able to filter the information available to connected vehicles from a multitude of connected device and to return to the driver only the information that is relevant to him/her. In order to illustrate the key features of our system, we used the parsing engine to design a distributed congestion management system. Validation was performed both via a microscopic traffic simulator and via a HIL platform. In order to validate our system via HIL testing, we designed new capabilities for the platform, which allow the HIL to be interfaced with a network of IoT objects. In particular, we designed new modules that can interface the HIL simulation with Twitter\textsuperscript \textregistered feeds and used those feeds to generate events in the simulation. Both the simulation and HIL validation results showed that the engine can be effectively used to design a distributed traffic management system and that the proposed approach has the potential to be an effective solution to the problem of a road obstructions. Specifically, in the case of irregular obstructions, it was shown that vehicular flow along the obstructed link can be successfully regulated, and the
rest of the excess of vehicular load can be balanced along adjacent alternative routes. In the case of regular obstructions, it was shown that vehicles can be allocated the limited resource (i.e. the obstructed link) in a fair way over the time as a result of the predictability of the obstruction.
The inclusion of a in-car decision engine also allowed the congestion management system for the re-routing of vehicles in a time effective manner, this enabling the use of more alternative routes to the obstruction.

Future work includes the design of more detailed route prediction algorithms which also take into account additional information such as historical data from the current trip and/or schedule patterns of historical trips, so that the Link Controller can be less affected by disturbances caused by mistaken predictions.

\section*{Acknowledgment}

This work has been conducted within the ENABLE-S3 project that has received funding from the ECSEL Joint Undertaking under grant agreement No. 692455. This Joint Undertaking receives support from the European Union’s Horizon 2020 research and innovation program and Austria, Denmark, Germany, Finland, Czech Republic, Italy, Spain, Portugal, Poland, Ireland, Belgium, France, Netherlands, United Kingdom, Slovakia and Norway. This work was also supported in part by SFI grant 11/PI/1177. 


\bibliographystyle{IEEEtran}		
\bibliography{ParsingEng}


\vspace{-0.5cm}
\begin{IEEEbiography}[{\includegraphics[width=1in,height=1.25in,clip,keepaspectratio]{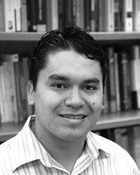}}]{Rodrigo~Ord\'{o}\~{n}ez-Hurtado}
received his Ph.D. degree from the University of Chile in 2012. He held a postdoctoral position at Maynooth University from 2012 to 2015 and at University College Dublin from 2015 to 2017, working with Professor R. Shorten and his research group on smart mobility applications and stability theory. In September, 2017 Rodrigo joined IBM Research - Ireland as Research Fellow. Rodrigo's interests include intelligent transportation systems with applications to smart cities, robust adaptive systems (control and identification), stability of switched systems, swarm intelligence, and large-scale systems.
\end{IEEEbiography}

\begin{IEEEbiography}[{\includegraphics[width=1in,height=1.25in,clip,keepaspectratio]{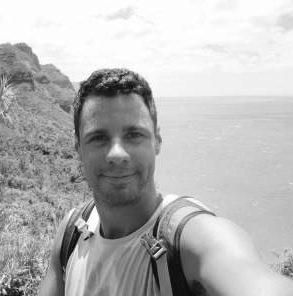}}]{Giovanni~Russo}
received his Ph.D. degree from the University of Naples Federico II in 2010. His focus was on the stability of nonlinear dynamical systems with applications to networked control and systems biology. Motivated by a desire to apply his results to real-world systems, soon after receiving his Ph.D. degree, he started to focus on automatic transportation systems. From 2012 to 2015, he was the lead system engineer and integrator of the Honolulu Rail Transit Project. In 2015, he joined IBM Research - Ireland. He is currently a member of the Board of Editors of IEEE TRANSACTIONS ON CIRCUITS AND SYSTEMS-I and of the IEEE TRANSACTIONS ON CONTROL OF NETWORK SYSTEMS.
\end{IEEEbiography}

\begin{IEEEbiography}[{\includegraphics[width=1in,height=1.25in,clip,keepaspectratio]{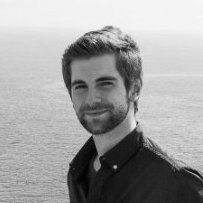}}]{Sam~Sinnott}
received his M.E degree in Electronic and Computer Engineering from University College Dublin, Ireland, in 2016. For his M.E. degree, he focused on the design of a congestion management system for single link congestion problems under the supervision of Prof. R Shorten.
\end{IEEEbiography}

\begin{IEEEbiography}[{\includegraphics[width=1in,height=1.25in,clip,keepaspectratio]{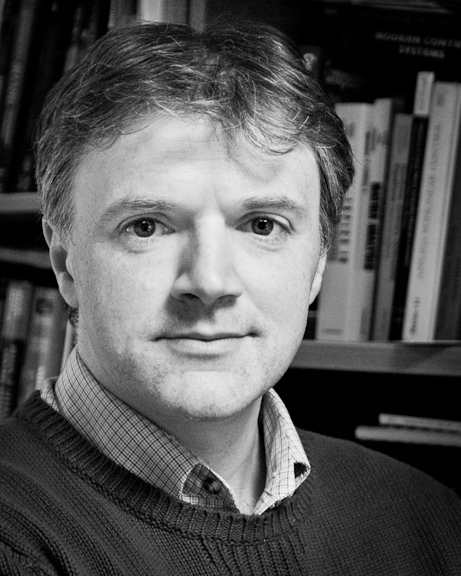}}]{Robert Shorten}
is currently Professor of Control Engineering and Decision Science at University College Dublin. Prof. Shorten's research spans a number of areas. He has been active in computer networking, automotive research, collaborative mobility (including smart transportation and electric vehicles), as well as basic control theory and linear algebra. His main field of theoretical research has been the study of hybrid dynamical systems.
\end{IEEEbiography}

\end{document}